\documentstyle[12pt]{article}

\newlength{\dinwidth}
\newlength{\dinmargin}
\newfont {\xx} {cmti10}

\newcommand{\nd}[1]{/\hspace{-0.6em} #1}

\def\be{\begin{equation}}
\def\l{\label}
\def\r{\ref}
\def\ee{\end{equation}}
\def\bea{\begin{eqnarray}}
\def\eea{\end{eqnarray}}
\def\nn{\nonumber \\}
\def \R{(\frac{ \alpha_i(0)}{ \alpha_i(t)})}
\def \RI{(\frac{\alpha_i(t)}{\alpha_i(0)})}
\def \R3{(\frac{\alpha_3(0)}{\alpha_3(t)})}
\def \RI3{(\frac{\alpha_3(t)}{\alpha_3(0)})}

\def \sc2e{\sum_iC_2(R_i)}
\def \sc2{$\sum_iC_2(R_i)$}
\def\c2e{C_2(R_i)}
\def\c2{$C_2(R_i)$}
\def\ta3{\tilde\alpha_3}

\def\b0{\frac{B}{M}}

\def\a2m{\frac{A^2}{M^2}}

\def\b0mu{\frac{B}{\mu}}

\def\a2mu{\frac{A^2}{\mu^2}}

\def\lapproxeq{\lower .7ex\hbox{$\;\stackrel{\textstyle <}{\sim}\;$}}
\def\gapproxeq{\lower .7ex\hbox{$\;\stackrel{\textstyle >}{\sim}\;$}}
\newcommand{\bean}{\begin{eqnarray*}}
\newcommand{\eean}{\end{eqnarray*}}

\begin{document}

\titlepage
\begin{flushright}
RAL-95-TR-057\\
December 1995 \\
\end{flushright}

\begin{center}
\vspace*{2cm}
{\Large {\bf Quark Model Description of Polarised Deep Inelastic
Scattering and the prediction of $g_2$}}
\\
\vspace*{1cm}
R.G.\ Roberts\\
Rutherford Appleton Laboratory, \\ Chilton, Didcot OX11 0QX, England.
\\
\vspace*{0.5cm}
 Graham G. Ross\\Department of Physics,
Theoretical Physics,\\
University of Oxford,
1 Keble Road,
Oxford OX1 3NP

\end{center}

\vspace*{2cm}
\begin{abstract}
We show how the operator product expansion evaluated in the
approximation of ignoring gluons leads to the covariant formulation
of the quark parton model. We discuss the connection with other
formulations and show how the free quark model prediction, $g_2=0$,
changes smoothly into the Wandzura-Wilczek (WW) relation for quark
masses small relative to the nucleon mass. Previous contradictory
parton model predictions are shown to follow from an inconsistent
treatment of the mass shell conditions.  The description is extended
to include quark mass corrections.

\end{abstract}

\newpage
\section {Introduction}

A recent preliminary measurement of the proton structure function
$g_2$\cite{slacg2}
in polarised deep inelastic scattering suggests that the integral
form of Wandzura-Wilczek\cite{ww,jrr} relation
\be
g_2(x) = -g_1(x) + \int_x^1 \frac{dx'}{x'} g_1(x')
\label{eq:1}
\ee
is consistent with the new data $-$ at least for x \gapproxeq 0.2.
Here $g_1$ is the structure function already measured with good
accuracy
from previous experiments\cite{slacg1,emc,smc}
with longitudinally polarised protons.
Polarising the proton in the transverse direction gives a measurement
of $g_1+g_2$ and hence an estimate for $g_2$ can be extracted from
both
sets of data.

In the operator product expansion (OPE) analysis of deep inelastic
polarised scattering both twist-2 and twist-3 operators contribute to
$g_2$ while only the former contribute to $g_1$. If one assumes the
twist-3 operator contributions are negligible then the moments of
$g_2$ are related to those of $g_1$ and, assuming the Mellin
transform can be performed\footnote{This may not be allowable due to
singular (Regge) behaviour at small $x$ in which case the moment
relations only may apply\cite{heimann}. In what follows we ignore
this potential problem.}, the relation of eq(\r{eq:1})
applies\cite{jrr}. The motivation for ignoring the twist-3 terms
follows from the observation\cite{sv} that, in the limit of zero
quark mass, all twist-3 operators involve gluons. Thus in the quark
model, with massless quarks and where gluons are neglected, one may
expect the WW relations to apply because the twist-3 operator matrix
elements vanish.   In a previous paper\cite{jrr}
(JRR) we showed  that a consistent covariant
formulation of the quark parton model, in which the massless on-shell
quarks
have a non-zero transverse momentum $k_T$, leads to precisely the WW
relation eq(\ref{eq:1}). However this formulation has been questioned
\cite{jaffe,jaffeji,ael} because it is apparently in contradiction
with other formulations of the parton model.

Because of this and motivated also by the encouraging results of the
preliminary SLAC
data\cite{slacg2} which lends some support to the quark model
approximation, we reconsider the determination of the polarised
structure functions in the
quark model approximation i.e. neglecting the gluonic component of
the nucleon but allowing for valence and sea quarks. The connection
with the OPE is  discussed in detail and we  show that the assumption
of negligible gluon component in the contributing operators is   
equivalent to a covariant parton
model description of the process in  which the quarks, which may be
massive, are allowed to have a completely  general transverse
momentum distribution. We find the covariant quark parton model
provides an entirely consistent picture of polarised scattering in
which the form of $g_2$ is fully determined once $g_1$ is measured.
The origin of previous contradictions between quark model relations
is shown to be due to incorrect treatment of the equations of motion
in the associated OPE.
 
We go on to consider the inclusion of quark masses in the parton model. 
In this case the parton polarisation vector picks up a component in
addition to the longitudinal one and the WW relation is no
longer satisfied. There is an additional polarised parton 
density associated with this component and restoring a connection 
between $g_1$
and $g_2$ rests upon relating the parton densities associated with
the two components. By considering plausible models for the source of
the parton polarisation, we show how such a relation can be made, 
leading once again to a prediction of $g_2$ in terms of the measured 
$g_1$.

\section {Covariant parton model for on-shell massless partons}

We start with a brief review of the covariant parton model
formulation of polarised scattering (in the massless limit) which we
shall show is equivalent to the OPE in the quark model approximation.
The anti-symmetric hadronic tensor is written in terms of the two
spin
dependent structure functions $g_1$ and $g_2$,
\be
W^A_{\mu\nu} (q,p,S) =i \varepsilon_{\mu\nu\rho\sigma}
\;\frac{q^\rho}{\nu}
\;[g_1(x) S^\sigma +g_2(x)  (S^\sigma -p^\sigma \frac{S.q}{M\nu})]
\label{eq:2}
\ee
where $M,p,S$ are the mass, momentum and polarisation vector of the
proton;
$q$ is the momentum of the virtual photon and $\nu = p.q/M$.
The covariant parton model expresses $W^A_{\mu\nu}$ in terms of a
convolution over the struck parton's momentum.
\be
W^{(A)}_{\mu\nu} (q,p,S) =\sum_h \int d^4k \;f_h (p,k,S)
\;w^A_{\mu\nu}
(q,k)\; \delta [(k+q)^2-m^2]
\label{eq:3}
\ee
Here $k,m,h$ are the parton's momentum, mass and (in the massless
limit) helicity ($h=\pm 1$)
and
\be
w^A_{\mu\nu} = ihm \varepsilon_{\mu\nu\rho\sigma} q^\rho s^\sigma
\label{eq:4}
\ee
where $s^\sigma$ is the parton spin vector which as $m^2 \rightarrow
0$ gives $m s_\sigma = k_\sigma$.
Summing over parton helicities thus
gives the combination of parton densities
$f_+-f_-\equiv \Delta f(p,k,S)$ and since we insist on
covariance, this can be written as $\Delta f(p.k,k.S,k^2)$.
As this combination has to be linear in the proton polarisation
$S_\sigma$
(spacelike),  covariance demands that $\Delta f \propto
k.S$ and so we can write
\be
\Delta f(p.k,k.S,k^2) =-\frac{(k.S)}{M} \tilde{f} (p.k,k^2)
\label{eq:5}
\ee
Substituting into eq(\ref{eq:3}), comparing with eq(\ref{eq:2}) and
using $d^4k \delta[(k+q)^2-m^2] = (\pi M/4 \nu) dk^2 dy$
where $y = 2\frac{p \cdot k}{M^2} = x + \frac{k_T^2}{xM^2}$
gives (for details see JRR)
\be
g_1(x) =\frac{\pi M^2x}{8} \int^1_x dy \; (2x-y) \; \tilde h(y)
\label{eq:6}
\ee
\be
g_2(x)=\frac{\pi M^2x}{8} \int^1_x dy \; (2y-3x) \; \tilde h(y)
\label{eq:7}
\ee
where $\tilde h(y) =\int dk^2 \tilde {f} (y,k^2)$. Crucial to
deriving
expressions (\ref{eq:6},\ref{eq:7}) and to the consistency of the
parton
model is the retention of the parton transverse momentum $k_T$. In
fact,
from (\ref{eq:6},\ref{eq:7}) we see that the combination
$g_1(x)+g_2(x)$
(which is the relevant quantity for a transversely polarised proton)
is proportional to $(y-x)$ which is simply $k^2_T/xM^2$. There is
just
{\it one} function $\tilde h(y)$ (the relevant charge weighted
combination
of parton distributions) which determines both $g_1$  and $g_2$
which is the key to obtaining a simple relation between the two.
Differentiating the sum $g_1 + g_2$ given by (\ref{eq:6},\ref{eq:7})
gives the relation eq(\ref{eq:1}).
This can be alternatively expressed in terms of moments \cite{ww}
\be
\int^1_0 dx\; x^{n-1} \;\left [ \frac{n-1}{n} g_1(x) +g_2(x)\right
]\;
=0
\label{eq:8}
\ee
the Burkhardt-Cottingham\cite{bc}(BC) sum rule being the $n=1$
version
\be
\int^1_0 dx\; g_2(x) = 0
\label{eq:9}
\ee

In section 4, we shall consider the corrections arising from allowing the
quarks to have a non-zero mass. 
These corrections lead to a violation of the WW sum rules eq(\ref{eq:8})
but making plausible assumptions for the origin of the parton polarisation,
$g_2$ can still be related to $g_1$. The BC sum rule eq(\ref{eq:9}) is
preserved exactly however. For a particular model the magnitude of the
violation of the WW sum rules can provide a possible phenomenological
estimate for the light quark mass.

\section{Consistency of the covariant quark parton model within the
OPE}

At a more formal level the polarised deep inelastic scattering may be
analysed using the OPE. Of course any viable model must be consistent
with the OPE and here we show that the covariant parton model is
equivalent to the OPE approach in the assumption of neglecting the
gluon component of the nucleon.
Let us start with the operator product expansion description of
polarised scattering
\bea
J_{\mu}(\xi)J_{\nu}(0)\mid_A=
i\epsilon_{\mu\nu\lambda\sigma}\partial^{\lambda}
\sum_{n=0,2,...}\left[ \sum_{i}F^n_{2,i}(\xi^2-i\epsilon
\xi_0,\mu^2)O_{2,i}^{\sigma\mu_1...\mu_n}(\mu^2)\right ]\nn
+\sum_{i}\left .
F_{3,i}^n(\xi^2-i\epsilon\chi_0,\mu^2)O_{3,i}^{\sigma
\mu_1...\mu_n}(\mu^2)\right ]\xi_{\mu_1}...\xi_{\mu_n}
\l{eq:sd}
\eea
where the twist-2 operators are given by
\be
O_{2,k}^{\sigma\mu_1...\mu_n}=
i^nS_1{\bar\psi(0)\gamma^{\sigma}\gamma^5 D^{\mu_1}...D^{\mu_n}
\lambda_k\psi(0)}-(Traces)
\l{eq:O2}
\ee
with $\lambda_{k=1..8}$ the $SU(3)$ Gell-Mann matrices and
$D^{\mu}=\partial^{\mu} +igA^{\mu}$ is the covariant derivative.
$S_1$
denotes symmeterisation in the $\sigma\mu_1...\mu_n$ indices. In
addition there is a flavour singlet operator corresponding to the
replacement of $\lambda_k$ by the unit operator together with a
flavour
singlet operator involving gluon fields only that we do not give
here.
The twist-3 operators are given by
\be
O_{3,k}^{\sigma\mu_1...\mu_n}=
i^nS A{\bar\psi(0)\gamma^{\sigma}\gamma^5 D^{\mu_1}...D^{\mu_n}
\lambda_k\psi(0)}-(traces)
\l{eq:O3}
\ee
where S denotes symmeterisation of the $\mu_1...\mu_n$ indices and A
anti-symmeterisation of the $\sigma \mu_i$ indices.  These operators
have a (suppressed) renormalisation scale ($\mu^2$) dependence. The
operator matrix elements are given by
\bea
<p,S|O_{2,k}|p,S>=
\frac{a_n}{n+1}S_1(S_{\sigma}p_{\mu_1}...p_{\mu_n}-traces)
\nn
<p,S|O_{3,k}|p,S>=
\frac{d_n}{n+1}S A(S_{\sigma}p_{\mu_1}...p_{\mu_n}-traces)
\eea
Using this in eq(\ref{eq:sd}) gives
\bea
\int_0^1dx \; x^n\; g_1(x,Q^2)=\frac{1}{4}a_n, \; n=0,2,4...\nn
\int_0^1dx\; x^n g_2(x,Q^2)=\frac{n}{4(n+1)}(d_n-a_n),\; n=2,4,...
\l{eq:oper}
\eea

In the case the twist-2 operators dominate, $d_n=0$ leaving just one
unknown matrix element per moment to describe two structure
functions. Thus one obtains the
WW relations eq(\ref{eq:8}) for $g_2$. Here we wish to demonstrate
that these relations follow in the quark model simply by analysing
the implications of the OPE under the assumption of negligible gluon
content of the nucleon. Of course there is no guarantee
that this assumption is realistic but it is of interest to determine
its
phenomenological implications so that experiment may determine
whether
a pure quark model picture ever makes sense. Further, having a quark  
parton model (albeit equivalent to the OPE) is useful in guiding one's  
intuition when discussing the long-distance effects as we discuss  
below. One thing to note is
that, due to QCD corrections, the pure quark model approximation can
make sense only for one value of $Q^2$ because the QCD corrections
that give perturbatively reliable information about the evolution in
$Q^2$ require that the gluonic component must become significant at
high $Q^2$ even if the quark picture is adequate at low $Q^2$. However 
if the pure quark model is a good approximation at one scale then one 
may conclude that the leading twist operators dominate at all higher 
scales, the gluonic component generated during
evolution being entirely leading twist in this case.

\begin{figure}[htb]
\vspace{7cm}
\includegraphics{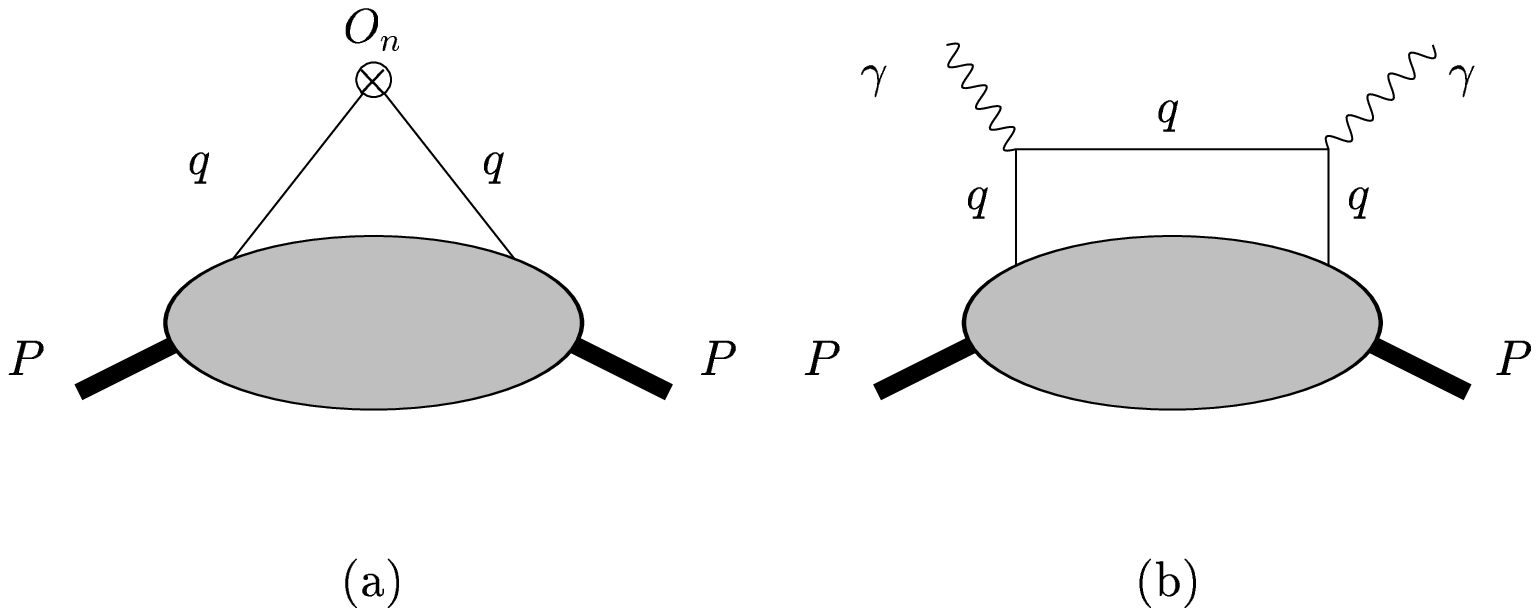}
\vspace{-1cm}
\caption{{\small a) Form of the operator matrix elements in the 
absence of gluons.
b) Form of the structure function using the OPE and neglecting the  
gluons both in the coefficient functions and the operator matrix  
elements. }}
\label{fig:1}
\end{figure}

Given this motivation we now consider the form of the quark (and
anti-quark) operator
matrix elements corresponding to Fig.1a. The contributing
operators
in this approximation are simply determined using the quark model
to determine the coefficients $F_{2,3}$ of eq(\ref{eq:sd}). Using
this, the sum over operators with the operator matrix elements
determined in the approximation of neglecting gluons is equivalent to
evaluating the diagram in Fig.1b. The latter requires
knowlege of the distribution $f_s(p,S,k,s)\equiv
f_s(p^2,k^2,p.k,k.S,p.s)$ of quarks with
polarisation $s$ within a proton. In the next section we will specify
these polarisation states precisely but here we wish to continue our
discussion within the framework of the OPE. The quark distributions
$f_s(p^2,k^2,p.k,k.S,p.s)$ in fact provide a convenient summary of
the effects
of the operator matrix elements and provide the easiest way of
determining the implications of the valence approximation for the
general OPE. Let us note the following facts concerning the OPE:

Operators related by equations of motion are not independent. In fact
Shuryak and
Vainstein\cite{sv} used
the equations of motion to show that, in the absence of quark masses,
the twist-3 operators introduced above are equivalent to operators
involving gluons only. Their general result (for massless quarks) is
\bea
O_{3,k}&=& S[ \frac{g}{8}\sum_{l=0}^{n-2}i^{n-2}\bar\psi D_{\mu_1}...
D_{\mu_{l}}\tilde G_{\sigma\mu_{l+1}}D_{\mu_{l+2}}
...D_{\mu_{n-1}}\gamma_{\mu_{n}}\lambda_k \psi \nn
\nn
&+&\frac{g}{16}\sum_{l=0}^{n-3}i^{n-3} \bar\psi D_{\mu_1} ...
D_{\mu_l}
(D_{\mu_{l+1}}\tilde G_{\sigma \mu_{l+2}})D_{\mu_{l+3}}...
D_{\mu_{n-1}}\gamma_{\mu_{n}}\lambda_k \psi  ]
\l{eq:sy}
\eea
where $\tilde G_{\alpha\beta}=\frac{1}{2}\epsilon
_{\alpha\beta\lambda\sigma}G^{\lambda\sigma}$.

Thus the twist-3 operator matrix elements will vanish in the valence
quark approximation up to mass effects. We shall discuss the massive
case shortly; in any case we may use the equations of motion to set
$k^2=m^2$, the quark mass squared.

The second point is that the normalisation scale dependence of the
operators is cancelled in the prediction for a physical quantity by
the
scale dependence of the coefficient functions $F_{2,3}$. Thus the
evaluation of Fig.1b which corresponds to the full observable
cross section has no scale dependence.

Putting these points together we see that the quark distributions in
fact are functions of just two variables
$f_s(p^2,k^2,p.k,k.S,p.s)=f_s(p.k, k.S)$
since $p^2$ and $k^2$ are given by the nucleon mass squared and the
quark mass squared respectively and, in the relativistic quark limit,
$p.s=p.k$.

Let us discuss the implications for cases of increasing complexity.
First consider the trivial case of scattering off a free on-shell
quark
moving collinearly with the target. In this case the quark
distribution $f_s$ is known (trivial) and the scattering amplitude is
simply given by the $\gamma q$ Born  diagram.  Some care must be
exercised in the massless limit.  Since $g_2$ is to be determined via
eq(\ref{eq:2}) (in this case  $p^{\mu}$
and $S^{\mu}$ are the {\it quark} momentum and spin) we must use a  
massive
quark
with mass $m$ otherwise the factor multiplying $g_2(x)$ vanishes. The
massless quark limit is then easily obtained by taking the limit
$m\rightarrow 0$. The Born diagram gives an amplitude proportional to
\be
im\epsilon_{\mu\nu\rho\sigma}q^{\rho}s^{\sigma}\delta((k+q)^2)
\ee
Comparison with eq(\ref{eq:2}) immediately shows that $g_2=0$. How
does this
relate to the OPE analysis? The operators contributing simply
correspond to those in the structure
\be
\frac{1}{((p+q)^2+m^2)}\bar \psi \gamma_{\mu} (\nd p +\nd q +m)
\gamma_{\nu} \psi
\l{eq:st}
\ee
The term $\psi \gamma_{\mu} (\nd q) \gamma_{\nu}
\psi\equiv i\epsilon_{\mu\alpha\nu\sigma}q^{\alpha} \bar \psi\gamma_5
\gamma^{\sigma}\psi$
when combined with the factors found expanding $(p+q)^{-2}=\sum_i
(-2p.q/q^2)^i/q^2$ and Fourier transforming to change the $p_{\mu}$
to
a derivative acting on the quark field, immediately leads to the
combination of operators
$O_{2}^{\sigma\mu_1...\mu_n}+O_{3}^{\sigma\mu_1...\mu_n}$. The reason
this combination arises is simply because there is no symmeterisation
(or antisymmeterisation) in the Born term between the $\sigma$ index
and
the indices associated with the $(p.q)^i$ terms. As discussed above
$O_2$ alone gives rise to the WW relations and
non-vanishing $g_2$. Thus the vanishing of $g_2$ must come about
through a cancellation of the contributions of $O_2$
and $O_3$. At first sight this seems impossible because of the
Shuryak
Vainstein relation eq(15) which suggests $O_3$ vanishes in the
absence of gluons. However here this is not true as a careful
computation of
the mass effects reveals. To illustrate the point consider the
simplest
operators
\bea
O_{2}= i^2S_1\bar \psi \gamma_{\sigma} \gamma_5 \partial_{\mu_1}
\partial_{\mu_2} \psi \nn
O_{3}= i^2SA\bar \psi \gamma_{\sigma} \gamma_5 \partial_{\mu_1}
\partial_{\mu_2} \psi
\eea
By writing $\partial _{\mu}=\{\gamma_{\mu},\nd \partial \}/2$ we may
re-express these as
\bea
O_2=i^2S_1\bar\psi \gamma_{\sigma}\nd \partial \gamma_{\mu_1}\gamma_5
\partial_{\mu_2}\psi \nn
O_3=imSA\bar\psi \gamma_{\sigma}\gamma_5\gamma_{\mu_1}
\partial_{\mu_2}\psi
\eea
We now take matrix elements of these operators between on-shell
partons
with polarisation vector $S_{\mu}$. In momentum space we have

\bea
<p,S|O_2|p,S> &\propto& Tr((\nd p+m)(1+\gamma_5\nd S)
\gamma_{\sigma}\gamma_5\nd p \gamma_{\mu_1})p_{\mu_2}\nn
&=&4m(S_{\sigma}p_{\mu_1}+S_{\mu_1}p_{\sigma})p_{\mu_2}
\nn
<p,S|O_3|p,S> &\propto& Tr((\nd p+m)(1+\gamma_5\nd S)
\gamma_{\sigma}\gamma_5
\gamma_{\mu_1})p_{\mu_2}\nn
&=&4m(S_{\sigma}p_{\mu_1}-S_{\mu_1}p_{\sigma})p_{\mu_2}
\eea
Thus we see explicitly that for the combination $(O_2+O_3)$ there is
a cancellation between $O_2$ and
$O_3$ of the term $\propto p_{\sigma}$ which is associated
with $g_2$. This cancellation persists in the massless limit. The reason
is that although the $O_3$
(and $O_2$) matrix element is formally of $O(m)$ this is cancelled by
the $1/m$ behaviour of the polarisation vector $s_{\mu}\sim
p_{\mu}/m$.
Thus $O_3$ {\it does} contribute even in the limit gluons are
ignored. We have somewhat belaboured our discussion of this very
simple model because this point has been the source of considerable
confusion in the literature.

So far our discussion has shown that $g_2$ vanishes in the free quark
model, a well known result. Of course a model of free collinear
quarks is not realistic and to determine the implications of more
reasonable models we consider now the case of Fig.1b with
nontrivial $f_s$. As we have stressed
the calculation of the full diagram is completely equivalent to the
use
of the OPE with operator matrix elements determined in the valence
approximation. Following the analysis presented above the result is
given in eqs.(\ref{eq:6}) and (\ref{eq:7}) and satisfies the WW
relation.  To understand the origin of this relation we note that the
intermediate (vertical) states of Fig.1b are on-shell quarks
and the result of eqs.(\r{eq:6}) and (\r{eq:7}) was derived to leading
order only in the quark
masses. To this order the quark polarisation vector is $s_{\mu}\sim
k_{\mu}/m$ and the matrix element of $O_3$ between the quark states
vanishes simply because one cannot form an antisymmetric tensor from
$k_{\mu}$ alone. This immediately implies that only $O_2$
contributes giving rise to the WW relation. This result again is
puzzling because there is no obvious
limit in which the free quark result with vanishing $g_2$ can be
obtained i.e. why doesn't
this argument apply too in the free quark case discussed above? The
answer is that the determination of $g_2$ requires the identification
of the coefficient of the second tensor in eq(\r{eq:2}). As we noted
above
this vanishes identically in the free quark limit (the approximation
that $S_{\mu}\sim p_{\mu}/m$) and for this reason we had to compute
$g_2$ in the massive quark case and take the massless limit
only at the very end. In the present case however the tensor
structure
of eq(\r{eq:2}) refers to the real nucleon and therefore doesn't
vanish
when using the approximation $s_{\mu}\sim k_{\mu}/m$. Thus the
argument
showing the vanishing of the $O_3$ contribution applies in this case
and we do indeed get the WW relation. Of course in the
limit the nucleon mass is taken small and the {\it nucleon}
polarisation satisfies  $S_{\mu}\sim p_{\mu}/M$ the
argument that $O_3$ vanishes fails and the WW relation
would not apply. This is the limit that establishes the connection
with
the free quark model result for it leads to $g_2=0$ again. Thus we
may
see the WW result applies in the limit $m/M$ and
$m/k_T$ are small while the free quark model result applies when
$m\sim M$, with $m/k_T$ small. In the next section we extend the
calculation to include mass effects to approach the
region where $m/M$ and $m/k_T$ are not negligible.

Before leaving the OPE analysis we should comment on the role of
another twist-3 operator not included in the above analysis. This is
the operator
\be
O_m=m S A\bar \psi
\gamma_{\sigma}\gamma_{\nu}\partial_{\mu_1}...\partial_{\mu_n}\psi
\ee
{}From eq(\r{eq:st}) we see this operator arises through the term
proportional to $m$. However it is easy to check in the free quark
model case that it gives rise to a gauge variant term which is
cancelled by the term proportional to $\nd p$ in eq(\r{eq:st})which  
involves the
operators $O_{2,3}$ and which have a matrix element proportional to
$m$.

Our analysis of the OPE in the approximation of ignoring gluons has
shown that it is equivalent to a covariant formulation of the quark
parton model. As discussed in \cite{jrr} this leads to a completely
consistent description of both polarised and unpolarised scattering.
The analysis also shows why other parton model formulations have led
to contradictory results because the OPE requires the use of
equations of motion to relate the various operators. This is in
conflict with a parton model formulation in which the parton momentum
is a fraction, $x$, of the nucleon momentum because the parton mass
will then be $xM$, an $x$ dependent quantity not consistent with the
equations of motion of the underlying quark states. Thus in these
models contradictory results for polarised scattering are obtained
(see \cite{ael} for a discussion of these results) but being in
conflict with the OPE their predictions should be ignored. As a
result we see that there is no remaining theoretical conflict for the  
case of the covariant quark parton
model description of polarised scattering. Whether it provides a good
phenomenological description of polarised scattering is a question
for experiment.

\section{Corrections from finite quark masses }

Only for massless quarks is the parton polarisation vector $s^\sigma$
proportional to $k^\sigma$. In the $m \neq 0$ case the partons will,
in general, have three polarisation components. Let us choose as a 
convenient basis the vectors $n_i$ given by
\bea
n_1^\mu &=& \frac{1}{mN_1}[(p.k)k^\mu - m^2p^\mu] \nn
n_2^\mu &=& \frac{1}{N_2}\epsilon^{\mu\nu\rho\sigma}k_\nu S_\rho
p_\sigma \nn
n_3^\mu &=& \frac{1}{N_1N_2}\epsilon^{\mu\nu\rho\sigma}k_\nu
\epsilon_{\rho\alpha\beta\gamma}k^\alpha S^\beta p^\gamma p_\sigma
\nn
 &=& \frac{1}{N_1N_2}[M^2(k.S)k^\mu +N_1^2 S^\mu -(p.k)(k.S)p^\mu]
\label{eq:nidef}
\eea
where
\bea
N_1 &=& [(p.k)^2 - m^2M^2]^{1 \over 2} \nn
N_2 &=& -[(p.k)^2 -M^2(k.S)^2 - m^2M^2]^{1 \over 2} 
\label{eq:n12}
\eea

In the parton rest-frame these vectors form a set of orthogonal unit
3-vectors and, in any frame, satisy $k.n_i = 0$, $n_i.n_j = -\delta_
{ij}$. In the proton rest-frame (PRF), they are given by
\be
n_1=\frac{1}{m}(|\vec k|,k_0 \hat {\vec k}), \quad
\hat {\vec n}_2 = \frac{1}
{\sin \theta} (\hat {\vec k} \times \hat {\vec S}), 
\quad \hat {\vec n}_3 =
\frac{1}{\sin \theta}[\hat {\vec k} \times (\hat {\vec k
}\times \hat {\vec S})]
\label{eq:niprf}
\ee
where $\hat {\vec k}.\hat {\vec S}=\cos \theta$ and $ 
\hat {\vec k}, \hat {\vec S}
$ are unit vectors in the 
direction of the parton momentum and proton spin respectively. The
charge weighted combination of the relevant parton distributions in
each direction we call $\Delta f_i(p,k,s)$. The vector $n_1$ is the
longitudinal polarisation vector of the parton, 
$mn_1^\mu = k^\mu +O(m^2/M^2)$ and so $\Delta f_1(p,k,s)$ is to
be associated with the $\Delta f(p,k,s)$ of eq(\ref{eq:5}). 

These $O(m^2/M^2)$ corrections modify the integrand of eq(\ref{eq:6})
and since the value of $k_T^2 = M^2[x(y-x)-m^2/M^2]$ the integrand 
of eq(\ref{eq:7}) is also appropriately modified.
In addition, there are $O(m^2/M^2)$ corrections to the limits of
the integration in $y$. If we write $\alpha = m^2/M^2$ then $y_{max} = 
1+\alpha$ follows from constraining $(p-k)^2 \ge 0$ 
\footnote{ Increasing this threshold to positive values of $O(m^2)$
leads to only minor changes since then $x \ge \alpha + O(\alpha^2)$} 
while $y_{min} = 
x + \alpha/x$ follows from $k_T^2 \ge 0$. The resulting expressions for
$g_1$ and $g_2$ associated with the component $n_1^\mu$ are 
\bea
g_1^{(1)}(x) &=& \frac{\pi M^2}{8} \int^{y_{max}}_{y_{min}} dy \;x(2x-y) \; 
(1-\frac{2\alpha}{xy}) \; \tilde h_1(y) \nn
g_2^{(1)}(x) &=& \frac{\pi M^2}{8} \int^{y_{max}}_{y_{min}} dy \; 
[x(2y-3x)+\alpha(4\frac{x}{y}-3)] \; \tilde h_1(y)
\label{eq:106}
\eea
Here, $\tilde h_1(y) = (1-\frac{4\alpha}{y^2})^{-\frac{1}{2}}
\int dk^2 \tilde f
(k^2,y)$ where $\Delta f_1(p,k,S) = -(k.S)/M \;\tilde f(k^2,y)$

Notice that $y_{max} \ge y_{min}$ only if $x \ge \alpha$, i.e. 
$g_{1,2}^{(1)}$
are zero for $x < m^2/M^2$. The above expressions do not satisfy the WW
sum rules(\ref{eq:8}) but $g_2^{(1)}$ does satisfy the 
BC sum rule(\ref{eq:9})
exactly. To see this we assume that we can interchange the order of the 
$x,y$ integrations and use the fact that the support of $x$, for fixed $y$,
is given by $x_{min} \le x \le x_{max}$ where $x_{min}$ and $x_{max}$ are
solutions of $x^2-xy+\alpha=0$.

The $n_2^\mu$ component is irrelevent since it is orthogonal 
to $S^\mu$ and so
$\Delta f_2(p,k,S)$  vanishes. The $n_3^\mu$ component does contribute however and substituting into
eqs(\ref{eq:3},\ref{eq:4}) gives the appropriate contributions to $g_1$
and $g_2$
\bea
g_1^{(3)}(x) &=& \pi mM^5\;\int^{y_{max}}_{y_{min}} dy 
\frac{x(y-x)-\alpha}{4N_1N_2} \int dk^2 \Delta f_3(p,k,s) \nn 
g_2^{(3)}(x) &=& \pi mM^5\;\int^{y_{max}}_{y_{min}} dy \frac{1}{8N_1N_2}
\left [ \frac{y^2}{2}-3x(y-x)+\alpha \right ] \int dk^2 \Delta f_3(p,k,s)
\label{eq:108}
\eea
where $y_{min}$, $y_{max}$ and $\alpha$ defined as above. The $g_1^{(3)}$
and $g_2^{(3)}$ defined by eq(\ref{eq:108}) satisfy analogous sum rules
to the WW sum rules in the limit $\alpha \longrightarrow 0$. Provided
$\Delta f_3 \propto N_2$, which is true in any reasonable model (see
below), we can 
write in this limit
\be
\int_0^1 dx x^{n-1} \left [ \frac{(n-1)(n-2)}{4n} g_1^{(3)}(x)
- g_2^{(3)}(x) \right ] = 0
\label{eq:109}
\ee
This relation is interesting in that not only is the BC sum rule satisfied
but also the first moment of $g_2^{(3)}$ vanishes
\footnote{It  
has been argued that the
same is true too for the contribution involving gluons  
\cite{ael,alt}.} which implies that $g_2^{(3)}$ is suppressed in general.
When $\alpha \neq 0$ the $n=1,2$ sum rules both survive exactly.

The above expressions, eq(\r{eq:106}) and eq(\r{eq:108}), are quite general. 
As they involve two distribution functions, they do not lead to a relation 
between $g_1$ and $g_2$. However, as we now discuss, the covariant parton 
model allows us to develop a plausible model 
to describe the source of the parton polarisation in the proton and hence to
relate $\Delta f_1(p,k,s)$ and $\Delta f_3(p,k,s)$.  How are the partons 
polarised? Ultimately it is due to the interaction with the external magnetic 
field which causes the proton
spin, through the interaction of the proton magnetic moment, to align with 
the field. At the parton level, while they too will interact directly with 
the magnetic field, they will also have spin-spin interations which align the 
individual parton spins.
The latter interaction, being due to the strong force, may be expected to 
dominate the {\it relative } parton spin alignment which is what relates 
$\Delta f_1(p,k,s)$ and $\Delta f_3(p,k,s)$. Since, on average, any 
combination of parton spins is proportional to the proton spin we are led 
to a model in which
the relative magnitude of $\Delta f_i$ is due to the spin-spin interaction 
between the partons themselves, the interaction being proportional
to the scalar product of the individual parton spin with the proton spin. 
Thus we have
\be
\Delta f_i(p,k,S) = (n_i.S)\;m\;\hat f(p.k,k^2)
\label{eq:110}
\ee
From eq(\ref{eq:nidef}) we have
\bea
m\;(n_1.S) &=& \frac{(p.k)(k.S)}{N_1} \nn 
m\;(n_2.S) &=& 0  \nn 
m\;(n_3.S) &=& -m\;\frac{N_2}{N_1}
\label{eq:111}
\eea
We see that factoring out $(k.S)$ in eq(\ref{eq:5}) in the 
massless case is consistent with this model. The factor $m$ on the rhs
of eq(\ref{eq:110}) ensures the desired limit for the $n_1^\mu$
contribution as $m \longrightarrow 0$. 
We can now write $\Delta f_3$ in terms of $\tilde f$,
\be 
\Delta f_3(p,k,S) = \frac{m N_2}{M(p.k)}\; \tilde f(k^2,y)
\label{eq:112}
\ee 
and substitute into eqs(\ref{eq:108}) to get
\bea
g_1^{(3)}(x) &=& \pi m^2\;\int^{y_{max}}_{y_{min}} dy 
\left [ \frac{x(y-x)-\alpha}{y^2} \right ]\tilde h_1(y) \nn 
g_2^{(3)}(x) &=& \pi m^2\;\int^{y_{max}}_{y_{min}} dy \frac{1}{2y^2}
\left [ \frac{y^2}{2}-3x(y-x)+\alpha \right ] \tilde h_1(y)
\label{eq:113}
\eea
Thus $g_1(x) = g_1^{(1)}(x) + g_1^{(3)}(x) $ 
and $g_2(x) = g_2^{(1)}(x) + g_2^{(3)}(x) $ are determined by a {\it single}
function once again, as in the $m =0$ case. The resultant prediction for
$g_2(x)$ in terms of the measured $g_1(x)$  is
pursued in the next section. 

In leading order the final form of the quark mass effects consists of 
two pieces. The first which starts at $O(\frac{m^2}{M^2})$  comes simply 
from imposing parton kinematics  on the scattering process. The second, 
of $O(\frac{m}{M})$, is given in eq(\r{eq:108}) 
and comes from the transverse polarisation states. The question immediately 
arises whether it is the constituent or current mass that is relevant. 
At the level of the quark parton model itself this cannot be answered as 
it is the QCD interactions
that cause masses to ``run". However one may make an educated guess as 
to the most appropriate choice. The calculation of the parton model process 
requires, for gauge invariance, that the mass of the intermediate parton 
be the same as that of the initial
parton. The former should certainly be taken as the current-quark mass at 
the scale $Q^2$ since the large momentum involved in the deep inelastic 
scattering process flows through it. Hence the struck quark mass should also be
taken to be the current-quark mass.
As a result we expect the corrections of $O(\frac{m}{M}) $ in eq(\r{eq:108}) 
to relate to the current quark mass. The kinematic corrections however have 
two origins. The first, due to the $y_{min}$ cut and from the form of the 
parton momentum $k$, comes
from putting the struck parton on mass shell and again for consistency 
should be taken to be the current-quark mass. The second, the $y_{max}$ 
cut, comes from imposing parton kinematics on the intermediate states in 
the lower bob of Fig 1b. This must 
surely be the constituent mass as no large momenta are involved. However, 
in practice, the former effects are the most important and hence we expect 
the dominant mass effects to be assocated with the current quark mass
(in the phenomenological analysis 
given below we make no distinction between the various masses 
- consistent with the parton model interpretation - but, following 
this discussion, we expect the masses should be of current quark magnitude). 

\section{Phenomenological analysis}

\begin{figure}[htb]
\includegraphics{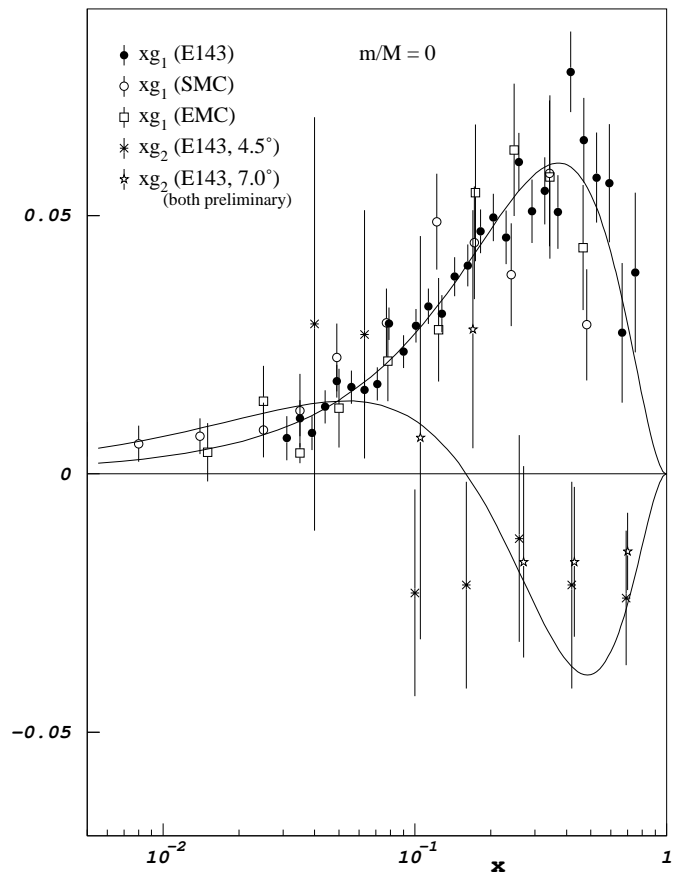}
\vspace{13cm}
\caption{ {\small Fit to the data on $g_1(x)$ from
refs(4,5,6) using eq.(6) and the comparison
of the resulting prediction for $g_2(x)$ from eq.(7) with
the preliminary data of ref(1). This is the $m=0$ case.
}} 
\label{fig:2}
\end{figure}

Fig.2 shows the present situation for the experimental measurements by
SLAC\cite{slacg1}, EMC\cite{emc} and SMC\cite{smc}   
for $g_1(x)$ together with the recent preliminary
measurements of $g_2(x)$ at SLAC\cite{slacg2}. 
The range of $Q^2$ over which all the measurements are taken is fairly  
wide and, in principle, one should
try to take account of possibly sizeable $Q^2$ variation at fixed $x$
values. However we do not tackle this here as we are concerned here only with the challenge of trying
to predict the size and shape of $g_2(x)$ from that of $g_1(x)$ at some
canonical value of $Q^2$ where the parton model may apply and, by definition,  QCD corrections are assumed to be small.

Also shown in Fig.2 is the expected $m=0$ prediction for $g_2(x)$ where
we fit to the data on $g_1(x)$ using the expression eq(\ref{eq:6})
with an assumed form for $\tilde h(y) =
a x^{-b}(1-x)^c$ which is then inserted into eq(\ref{eq:7}) to 
give $g_2(x)$. That is, the curve on Fig.2 is the prediction for
$g_2(x)$ given by the WW relation eq(\ref{eq:1}). We find $b \sim 3$
and $c \sim 1$. Certainly at large
$x$ it is encouraging to see the preliminary $g_2$ data lying close
to the WW expectation. Note that, from eq(\ref{eq:1}), the prediction
for $g_2(x)$ at any $x$ value depends only on values of $x$ at and  
above
that value and so is not sensitive to uncertainties in the small $x$
region.  Likewise, parametrisations  of the distributions $\tilde h(y)$
are not required to satisy various theoretical conjectures for the  
small
$x$ behaviour since this region is quite irrelevant to our concerns  
here.

\begin{figure}[hp]
\vspace{5cm}
\includegraphics{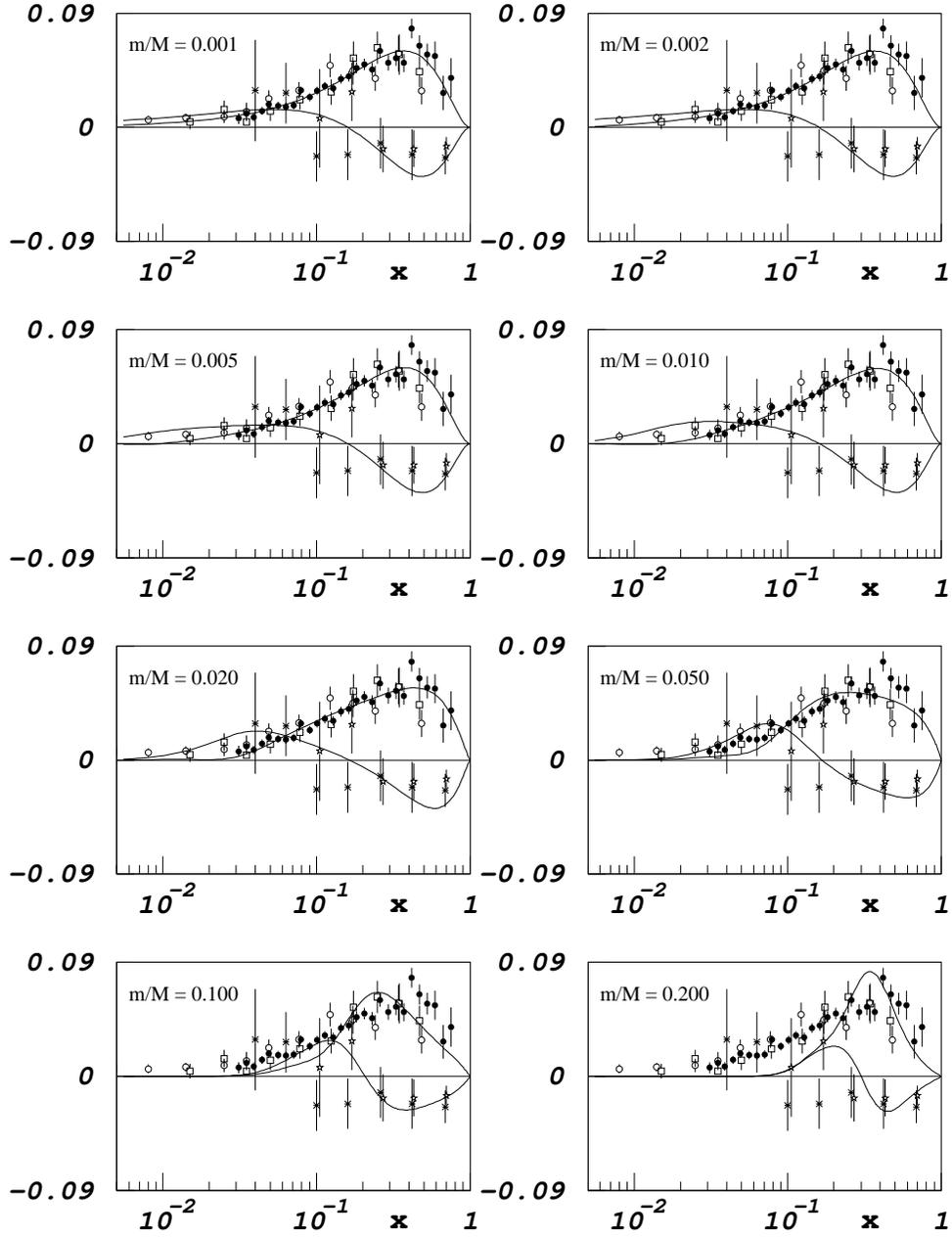}
\vspace{12cm}
\caption{{\small Fits to the data on $g_1(x)$ from
refs(4,5,6) using eqs.(25,31)
for $m/M$ up to 0.2  and the comparison
of the resulting predictions for $g_2(x)$ from
eqs.(25,31) with
the preliminary data of ref(1).}}
\label{fig:3}
\end{figure}

\begin{figure}[hp]
\vspace{5cm}
\includegraphics{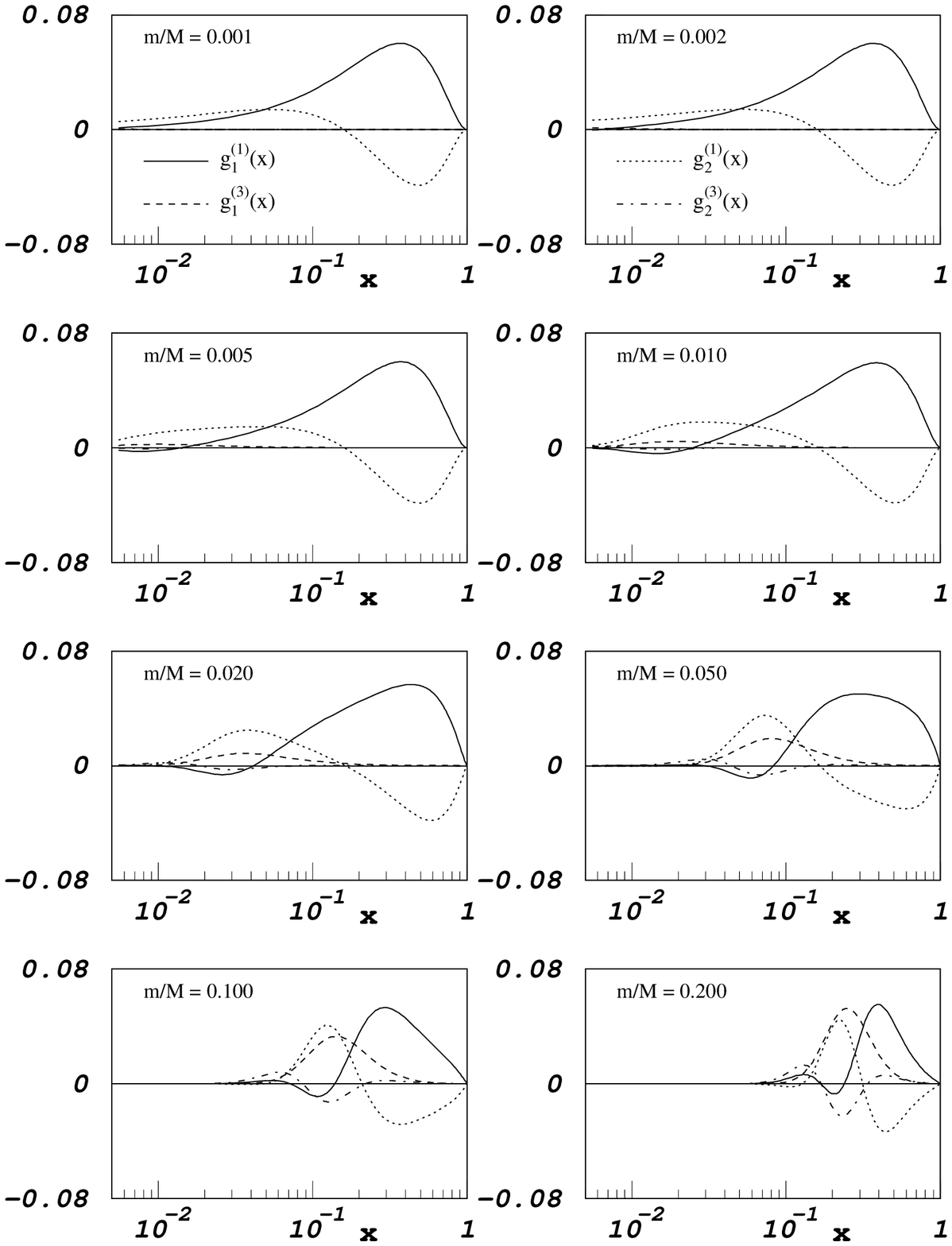}
\vspace{12cm}
\caption{ {\small The two components $g_i^{(1)}(x)$ and $g_i^{(3)}(x)$ 
for the fits shown in Fig.3
}} 
\label{fig:4}
\end{figure}

Our conclusion following from Fig.2 is that the preliminary $g_2$ data,  
albeit with relatively large errors, is consistent with WW relation and  
hence, 
within the framework of the covariant partom model, consistent with a
light quark mass of zero.  The next step is to ask if the data are 
also consistent with a sizeable quark mass. 

To answer this question, we carried out fits to $g_1(x)=g_1^{(1)}(x)
+ g_1^{(3)}(x)$ using
eqs.(\ref{eq:106},\ref{eq:113}) for $m/M$ up to 0.2 and a similar
parametrisation for $\tilde h(y)$ as in the $m=0$ case above.
The resulting fits, together with the corresponding predictions for
$g_2(x)=g_2^{(1)}(x) + g_2^{(3)}(x)$ are shown in Fig.3.
The quality of the fits is good provided $m/M \lapproxeq 0.04$, however
the resulting $\chi^2$ never {\it improves} on the value of 59 for
50 data points achieved by the $m=0$ fit. This remains true even when
more complicated parametrisations of $\tilde h(y)$ are considered.

The components of $g_1(x)$ and $g_2(x)$ associated with the $n_1^\mu$
and $n_3^\mu$ components of the parton polarisation are shown in 
Fig.4. As $m/M$ increases, $g_1^{(3)}(x)$ grows at low $x$ $-$ the
suppression of $O(m^2/M^2)$ is largely offset by the denominator
$y^2$ in the integrand $-$ and tends to spoil the quality of the fits.
Notice in Fig.4 that each component of $g_2(x)$ correctly integrates to
zero and note how $g_2^{(3)}(x)$ is suppressed relative to 
$g_2^{(1)}(x)$ due to the vanishing of the first moment.
 
To quantify the violation of the WW sum rules in this particular 
model for the quark mass effects we consider the ratio
$r_n$ given by
\be
r_n = \frac{\int^1_0 dx\; x^{n-1} \;\big [ \frac{n-1}{n} g_1(x)  
+g_2(x)\big ]}
{\int^1_0 dx\; x^{n-1} \;\big [ \frac{n-1}{n} g_1(x) \big ]}
\label{eq:20}
\ee
That is, $r_n$ is the WW moment normalised by the $g_1$ contribution
to that moment. Up to values of $m/M$ where the model is able to successfully
describe the data, $r_n$ can be used as a direct measure of the quark mass.
For values of $m/M = 0.01,\;0.02,\;0.04$ we get $r_2=1.2,\;3.4,\;9.0\% $.
   
As the precision of the SLAC $g_2$ measurement increases, we expect that
a phenomenological analysis such as this can offer a new and practical 
procedure for testing whether the quark model is a good approximation.

\medskip

\noindent
{\bf Acknowledgement }

\noindent
 We would like to thank G. Altarelli and K. Ellis for useful discussions.

\end{document}